# CCT, A CODE TO AUTOMATE THE DESIGN OF COUPLED CAVITIES *

P. D. Smith, General Atomics, Los Alamos, NM 87544, USA

*Abstract*

The CCT (Coupled Cavity Tuning) code automates the RF calculations and sizing of RF cavities for the CCL (Coupled Cavity Linac) structures of APT. It is planned to extend the code to the CCDTL (Coupled Cavity Drift Tube Linac). The CCT code controls the CCLFISH code, a member of the Poisson Superfish series of codes [1]. CCLFISH performs RF calculations and tunes the geometry of individual cavities, including an accelerating cavity (AC) and a coupling cavity (CC). CCT also relates the AC and CC by means of equations that describe the coupling slot between cavities. These equations account for the direct coupling, the next nearest neighbor coupling between adjacent AC's, and the frequency shifts in the AC and CC caused by the slot. Given design objectives of a coupling coefficient $k$, the pi/2 mode frequency, and the net frequency of the CC, the CCT code iterates to solve for the geometry of the AC and CC and the separation distance between them (this controls the slot size), satisfying the design objectives. The resulting geometry is used to automate CAD drawing preparation. The code can also be used in "as-designed" mode to calculate the frequencies and coupling of a specified configuration. An overview of the code is presented.

## 1 REASONS FOR DEVELOPMENT

The development of CCT was undertaken to speed up the design and tuning of coupled cavities in the Low Energy Linac of APT. The Low Energy Linac, which accelerates a cw beam of protons to 212 MeV, is subdivided into 341 segments, each of which has a unique cavity design (AC, CC, and slot).

The design objective is to specify the frequencies of the AC and CC and the coupling constant $k$ and to solve for the cavity geometry that achieves this. Each AC has a specified length and gap (defined by the Parmila code). The AC diameter is tuned using CCLFISH. Each CC has a specified length and diameter. The CC post gap is tuned using CCLFISH. The geometry of the slot between each AC-CC pair is a function of the unknown AC and CC dimensions and the separation distance between cavities. The slot geometry influences both the coupling coefficient $k$ and the net frequencies of the AC and CC. Therefore, to obtain a consistent design solution, it is necessary to iterate between CCLFISH tuning calculations and slot coupling calculations, as will be explained in more detail in section 2.4.

The design was formerly done by manually iterating between CCLFISH runs and slot calculations, using a spread sheet to control the process. If each segment were designed this way, up to 6 weeks per segment and 40 man years for the complete accelerator could be required. Fortunately, one can simplify this by designing fewer segments, tuning the cavities and the slot insertion with the aid of cold models, and interpolating the tuned geometry between segments, as is now standard practice.

With the CCT code, a lower cost approach can be used. Now, each design calculation can be performed in minutes rather than weeks, so it is possible to tune each segment analytically. Instead of interpolating the geometry, the approach is to interpolate empirical factors used in the coupling calculations. The empirical factors may be estimated fairly accurately with fewer, simpler cold models and the use of 3-D RF calculations to supplement the experiments [2].

Regardless of the strategy to be employed for cold models and interpolation, the CCT code has enabled a significant reduction in effort per design calculation, and it has permitted us to perform many more design calculations than would otherwise be possible, to investigate different design approaches and tuning.

## 2 DESCRIPTION OF THE CCT CODE

### 2.1 Geometry

CCT models half of an AC, half of a CC, and the coupling slot between them ( Fig. 1).

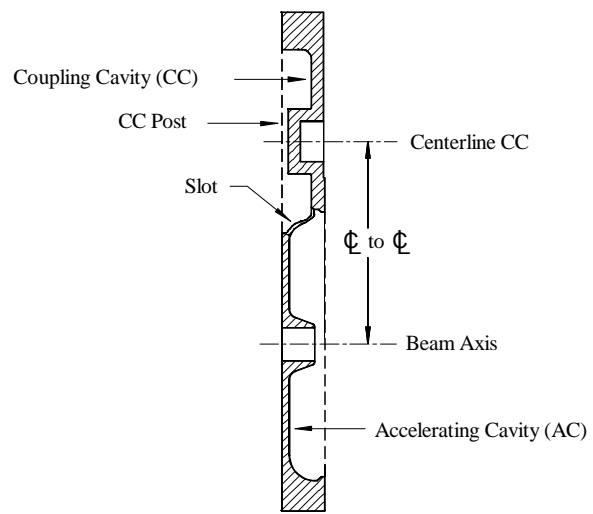

Figure 1: CCT cavity geometry

---

* Work supported by the APT project, U.S. DOE contract DE-AC04-96AL89607

The half cavities are nominally symmetrical, representing an infinite, biperiodic structure. Asymmetry can be introduced by an option to specify the frequency shift of the "opposite" coupling slot manually.

In the accelerator, the cavity plates (Fig. 1) are assembled with adjacent AC's back-to-back, creating a CC between them. The AC's communicate magnetically through their common CC, giving rise to "next-nearest-neighbor coupling" between them. This causes the $\pi/2$ mode frequency to be lower than the average AC frequency (see Eq. 2). Adjacent coupling cavities, however, are rotated 180° apart, so the next-nearest-neighbor coupling between two CC's is negligible.

The RF model used by CCLFISH (which is called by CCT) is axisymmetric, i.e., the slot is absent. The frequency shift and coupling effects of the slot are calculated separately by CCT. To do this, CCT must first calculate the geometry of the slot (length, width, and chamfer thickness) from the dimensions of the cavities and their center-to-center distance. Although the actual slot is a three-dimensional surface intersection, the key slot dimensions (length and width) can be calculated exactly using two-dimensional trigonometry. The chamfer thickness of the slot is calculated by an approximation that is corrected with the aid of a 3-D CAD model.

The slot geometry calculations are too complicated to present here in detail – several combinations of surface intersections are possible. Fig. 2 illustrates the geometry of the slot width calculation for one case.

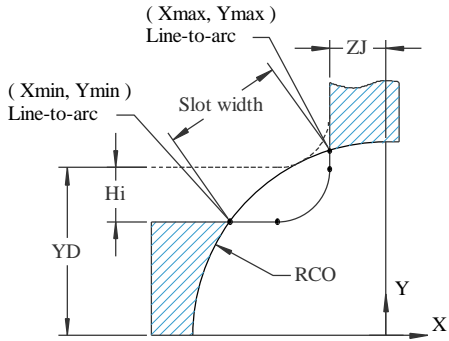

Figure 2: Example illustrating calculation of slot width

For the case shown in Fig. 2, the following equations are used to calculate the slot width:

$$X_{min} = -\sqrt{RCO^2 - (YD - H_i)^2}$$

$$Y_{min} = YD - H_i$$

$$X_{max} = -ZJ$$

$$Y_{max} = +\sqrt{RCO^2 - ZJ^2}$$

$$Slot\ width = \sqrt{(X_{max} - X_{min})^2 + (Y_{max} - Y_{min})^2}$$

## 2.2 Accounting for Frequency Shifts

The CCT code accounts for the following frequencies and frequency shifts.
1. The frequency calculated by CCLFISH in the absence of a coupling slot, $f_{SF}$.
2. A finite mesh correction term, $\Delta f_{mesh}$, subtracted from #1 to obtain the "real" frequency with no slot.
3. The frequency shift, $\Delta f_{slot}$, caused by the presence of (usually) two coupling slots in the cavity.
4. The net frequency, $f_{net}$, adjusted for the effect of the slot, given by

$$f_{net} = f_{SF} - \Delta f_{mesh} - \Delta f_{slot} \quad (1)$$

5. The accelerating mode $\pi/2$ mode frequency for the structure is the net frequency of the AC adjusted for the next-nearest-neighbor coupling, given by

$$f_{\pi/2} = \frac{f_{SF} - \Delta f_{mesh} - \Delta f_{slot}}{\sqrt{1 - kk}} \quad (2)$$

where $kk$ is the next-nearest-neighbor coupling coefficient. The value of $kk$ is negative.

The values of $f_{\pi/2}$ and $\Delta f_{mesh}$ are specified as input to the CCT code. $\Delta f_{slot}$ and $kk$ are found iteratively by CCT code using the solution logic explained in 2.4.

## 2.3 Slot Coupling Equations

The effects of the coupling slot are based on theoretical approaches presented by Gao [3] (for the frequency shifts and the direct coupling, $k$) and Greninger [4] (for the next nearest neighbor coupling, $kk$). These, in turn, are based on the Slater perturbation theory. The theoretical equations are adjusted by empirical "A" factors to account for the fact that the perturbation theory is not exact. Following is a summary of the equations used.

The coupling coefficient $k$ between an AC and a CC that are connected by an elliptical slot is given by:

$$k = A_k \left( k_{magnetic} + k_{elecctric} \right)$$

where

$A_k$ is an empirical correction factor,

$$k_{magnetic} = \frac{\pi}{3} m_0 \left(\frac{L}{2}\right)^3 \left(\frac{e_0^2}{K(e_0) - E(e_0)}\right) \frac{H_{ac} H_{cc}}{\sqrt{U_{ac} U_{cc}}} e^{-a_H t}$$

$$k_{electric} = -\frac{\pi}{3} e_0 \left(\frac{L}{2}\right)^3 \left(\frac{W}{L}\right)^2 \left(\frac{1}{E(e_0)}\right) \frac{E_{ac} E_{cc}}{\sqrt{U_{ac} U_{cc}}} e^{-a_E t} W$$

$W$, $L$, and $t$ are the full width, full length, and thickness respectively of the slot,
$E$ and $H$ are electric field magnetic field in a particular cavity evaluated at a location central to the slot,
$U$ is the stored energy in the cavity, and

$$e_0 = \sqrt{1 - \frac{W^2}{L^2}}.$$

$K(e_0)$ and $E(e_0)$ are complete elliptic integrals of the first and second kind, defined as follows:

$$K(e_0) \equiv \int_0^{\pi/2} \frac{d\varphi}{\sqrt{1-e_0^2 \sin^2\varphi}}$$

$$E(e_0) \equiv \int_0^{\pi/2} \sqrt{1-e_0^2 \sin^2\varphi}\, d\varphi, \text{ and}$$

$\alpha_E$ and $\alpha_E$ are decay rates for evanescent modes in an elliptical waveguide (details not given here).

The frequency shift for a particular cavity, either the AC or the CC, is given by

$$\Delta f = A_f \left( \Delta f_{magnetic} + \Delta f_{electric} \right)$$

where

$A_f$ is an empirical factor, different for the AC and CC,

$$\Delta f_{magnetic} = f_{no\_slot} \frac{\pi \mu_0}{12} \left(\frac{L}{2}\right)^3 \left(\frac{e_0^2}{K(e_0)-E(e_0)}\right) \frac{H^2}{U} e^{-\alpha_H t}$$

$$\Delta f_{electric} = f_{no\_slot} \left(\frac{\pi\, \varepsilon_0}{12\, E(e_0)}\right) \left(\frac{W}{L}\right)^2 \left(\frac{L}{2}\right)^3 \left(\frac{E^2}{U}\right) e^{-\alpha_E t}$$

$f_{no\_slot}$ refers to the CCLFISH frequency with correction for a finite mesh. Other variables are defined above. Values applicable to the AC or CC are used as appropriate.

The next-nearest-neighbor coupling coefficient $kk$ between two adjacent accelerating cavities is given by

$$kk = -A_{kk} \frac{\pi\, \mu_0}{18} \left( \frac{\left(\frac{L}{2}\right)^3 e_0^2}{K(e_0)+E(e_0)} \right)^2 \frac{H_{AC}^2}{|x|^3 U_{AC}} e^{-2\alpha_H t}$$

where $A_{kk}$ is an empirical factor, and $|x|$ is the distance between the centers of two adjacent slots in the CC.

## 2.4 Solution Logic

The logic used to obtain a consistent, coupled solution for both the AC and the CC is as follows:
1. Specify values for the cavity dimensions, including those that will later be adjusted by CCLFISH.
2. Assume values of the frequency shifts caused by the slots and the next-nearest-neighbor coupling.
3. Starting with the design objective π/2 mode frequency and CC net frequency, work backward through Eqs. (1) and (2) to calculate the shifted CCLFISH target frequencies, $f_{SF}$.
4. Run CCLFISH to tune the AC and CC to meet these shifted target frequencies. This modifies the cavity diameters or gaps.
5. Assume a center-to-center distance between the tuned cavities.
6. Given the geometries of the tuned cavities and the assumed center-to-center distance, calculate the geometry of the knife edge slot.
7. If the slot chamfer is specified in terms of a thickness on the "major diameter" of the slot, solve (by iteration) for the radial chamfer that will produce the specified thickness.
8. Apply the chamfer to the slot by increasing its length and width dimensions.
9. Given the calculated slot geometry and the fields and energies from CCLFISH, calculate the coupling coefficients $k$ and $kk$.
10. Compare the calculated $k$ with the design target $k$. Determine another guess for the center-to-center distance. Go to Step 6. Iterate on the center-to-center distance until $k$ meets its design objective.
11. Given the converged slot, calculate new frequency shifts $\Delta f_{slot,AC}$ and $\Delta f_{slot,CC}$. Go to Step 3. Iterate until the frequencies have converged.

# 3 EXPERIENCE USING CCT

The CCT code has greatly increased our productivity in designing coupled cavities. A CCL cold model that was designed using the code is shown in Fig. 3.

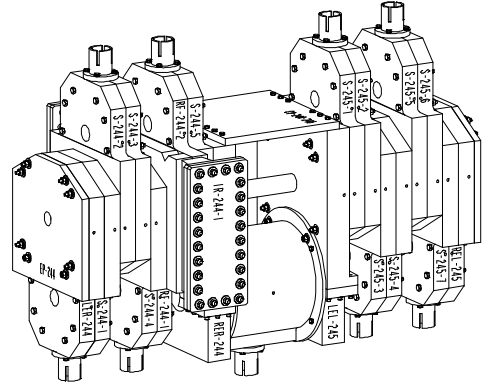

Figure 3: Example of cold model designed with CCT

Agreement between design and experiment has been good, although the perturbation theory has limitations. Improved accuracy is expected when we use 3-D analysis to refine estimates of the "A" factors.